\documentclass[12pt,preprint]{aastex}
\usepackage{psfig}
\def\HI {H\kern0.1em{\sc i}} 
\def\radm {rad m$^{-2}$}

\def\dg{$^{\circ}$}
\begin{document}
\title{Magnetic Fields in the Center of the Perseus Cluster}
\author{G. B. Taylor\altaffilmark{1,2}, N. E. Gugliucci\altaffilmark{2,3},
A. C. Fabian\altaffilmark{4}, 
J. S. Sanders\altaffilmark{4}, G. Gentile\altaffilmark{1}, and 
S. W. Allen\altaffilmark{5}}

\altaffiltext{1}{University of New Mexico, Dept. of Physics and 
Astronomy, Albuquerque, NM 87131, USA; gbtaylor@unm.edu}
\altaffiltext{2}{National Radio Astronomy Observatory, Socorro, NM
  87801, USA}
\altaffiltext{3}{Department of Astronomy, University of Virginia,
  Charlottesville, VA 22903; neg9j@virginia.edu}
\altaffiltext{4}{Institute of Astronomy, Madingley Road, Cambridge CB3 
  0HA, UK; jss, acf@ast.cam.ac.uk}
\altaffiltext{5}{Kavli Institute of Particle Astrophysics and Cosmology,
Stanford University, Stanford, CA 94305, USA; swa@stanford.edu}



\begin{abstract}

We present Very Long Baseline Array (VLBA) observations of the nucleus
of NGC\,1275, the central, dominant galaxy in the Perseus cluster
of galaxies.  These are the first observations to resolve the linearly
polarized emission from 3C\,84, and from them we determine a Faraday
rotation measure (RM) ranging from 6500 to 7500 \radm\ across the tip
of the bright southern jet component. At 22 GHz some polarization is
also detected from the central parsec of 3C\,84, indicating the
presence of even more extreme RMs that depolarize the core at lower
frequencies.  The nature of the Faraday screen is most consistent with
being produced by magnetic fields associated with the optical
filaments of ionized gas in the Perseus Cluster.

\end{abstract}


\section{Introduction}

The Perseus Cluster, A426, is the most X-ray luminous cluster in the nearby
universe, and the prototypical ``cooling core'' cluster.  In these
cooling core clusters the radiative cooling time of the X-ray emitting gas
is considerably shorter than the age of the universe, so to maintain
equilibrium the gas must flow into the center of the cluster, or
another source of energy is required to re-heat the gas.  Since
massive flows of material are not observed, the emerging solution to
the energy deficit is the active galaxy that is almost always \citep{bur90}
present at the center of massive clusters.  Shocks and ripples are
clearly evident in the deep Chandra image of Perseus \citep{fab05, fab06},
and could provide steady heating of the center of the cluster
\citep{fab05b}.  In Perseus the AGN manifests itself directly as a
bright radio source known as Perseus~A or 3C\,84, associated with
the galaxy NGC\,1275.  3C\,84 is one of
the brightest compact radio sources in the sky and has been studied in
some detail \citep{ver94, tay96, sil98, wal00}.  In particular it is
known to undergo bursts of activity \citep{kel68,tay96} that could
also drive the observed shocks and sound waves through the cluster.

Faraday rotation measure (RM) observations with the Very Large 
Array (VLA) of radio galaxies embedded
in clusters of galaxies have been used to elicit information about the
magnetic field strength and topology associated with the hot cluster
gas \citep{tay94, car02}.  These magnetic fields then play an
important role in modifying the energy transport and dissipation in
the center of the cluster \citep{fab05b}.  Comparisons between the
inferred cooling flow rates (in the absence of re-heating) and the
maximum Faraday rotation measure indicate a correlation, and there is
some evidence that magnetic fields are enhanced at the centers of
clusters \citep{fer99,vog05}.  Given the high density of gas at the
center of the Perseus cluster the Faraday rotation measures towards
3C\,84 are expected to be well over 1000 \radm\ \citep{tay94}.
Unfortunately such high RMs, and the
correspondingly large RM gradients are difficult to measure due to
cancellation of the linearly polarized signal within the telescope
beam (typically 0.5\arcsec for VLA observations).  Another problem
with obtaining polarimetric measurements of 3C\,84 with the VLA is the high dynamic range
imposed by the $\sim$20 Jy peak flux density at centimeter wavelengths
and arcsecond resolution.  In order to overcome these limitations we
have used simultaneous multi-frequency observations of 3C\,84 at high
angular resolution taken using the Very Long Baseline Array.

Throughout this paper we assume $H_0$ = 71 km s$^{-1}$ Mpc$^{-1}$ so that
1\arcsec\ = 0.35 kpc at the redshift of NGC\,1275 (0.0176; \cite{huc99}).

\section{VLBA Observations}\label{VLBAObs}

Observations were centered on 4.8~GHz, 8.4~GHz, 15.1~GHz, and 22.2~GHz
with the VLBA on 26 October 2004 and 11 November 2004 using the
VLBA\footnote {The National Radio Astronomy Observatory is operated by
Associated Universities, Inc., under cooperative agreement with the
National Science Foundation.}. In both cases, 3C\,84 was being used as
the leakage term calibrator for observations of Compact Symmetric
Objects \citep{gug05}.  Six scans of typically 2 minutes duration were
obtained at each frequency band.  Each frequency band was separated
into four IFs, and these IFs were paired for the purpose of imaging
the total intensity except at 22.2~GHz where all four IFs were
averaged.  For the purposes of determining the polarization and
subsequently the RMs each IF was imaged separately in Stokes Q
and U. Observational parameters are presented in Table 1.  

Amplitude calibration of the data was derived from system temperatures
and antenna gains.  Fringe-fitting was performed with the AIPS task
FRING on 3C~84.  The leakage, or D-term solutions were determined with
the AIPS task LPCAL on 3C~84.  Absolute electric vector position angle
(EVPA) calibration was determined using the EVPAs of J1310$+$322 and
BL~Lac listed in the VLA Monitoring
Program\footnote{http://www.vla.nrao.edu/astro/calib/polar/}
\citep{tmy00}.  Note that the EVPAs were corrected for each of the
four IFs separately.

\section{Chandra Observations}\label{XrayObs}

We used recent deep Chandra observations \citep{fab06} of the 
Perseus cluster to model the central density profile.  
In Fig.~1 we show the deprojected density profile and the 
projected temperature profile.  Within the central 0.8 kpc
(2.2\arcsec) the profile is severely affected by the 
nucleus so that the temperatures and densities are not
representative of the properties of the intracluster
medium (ICM).  We estimate an average central density over the
inner 2 kpc to be 0.3 cm$^{-3}$.

\section{Results}

We formed total intensity images of 3C\,84 between 5 and 22 GHz.
Images of similar quality are readily available in the literature
\citep{ver94, wal00}, so we do not reproduce them here.  Instead we
provide a brief summary of source properties (position, flux density,
size, distance, etc.) in Table 2, and we concentrate here on the 
polarimetry results.  

We formed linear polarization images at constant resolution (see
Fig.~2) for the purpose of comparing the polarization properties of
3C\,84 as a function of frequency.  The resolution in Fig.~2 is set by
the 5 GHz observations and the higher frequencies have been tapered to
provide matching resolution.  Linear polarization is detected on
November 11 from the bright jet component S1 in 3C\,84 at 5, 8, 15 and
22 GHz at a level of 0.8 to 7.5\% increasing with frequency.
Furthermore, there is some suggestion at 8.4 GHz and above that the
polarization is extended.  Similar results were obtained on October
26, but the leakage calibration was not as good at this epoch, so for
the remainder of the discussion we focus on the November 11 results.

The detection of core polarization is less than 0.1\% for all
frequencies except for 22 GHz for which it is 0.2\%.  At this level
the 14 mJy of linearly polarized flux density could well be
significant.  There is also some suggestion of polarization at 22 GHz
in between the core and the end of the bright inner emission
(component S1).

In Fig.~3 we present the RM image, a pixel-by-pixel fit to the
polarization angle as a function of the square of the wavelength
($\lambda^2$).  Since the 5 GHz observations were only weakly
polarized and did not resolve S1, we have not included them in the
fit.  The 8 frequencies included in the fit were: 8.114, 8.184, 8.421,
8.594, 14.906, 14.972, 15.269, 15.368, and 22.233 GHz.  The 15 and 22
GHz observations were tapered to match the 8.4 GHz resolution of 1.8
$\times$ 1.3 mas$^2$. Pixels were blanked if the statistical error in
polarization angle exceeded 5 degrees at any frequency.    A
representative fit at the peak of the polarized flux density of S1 is
shown in Fig.~4.  The systematic uncertainties in the polarization
angle measurements were assumed to be $\sim$3 degrees.

The rotation measure image (Fig.~3) shows a gradient of about 1000
\radm\ pc$^{-1}$ across component S1 in the southern jet component.
The statistical error in the RM determinations are $\sim$60 \radm, so
it is likely that this gradient is real.  It is possible that spectral
effects (see Fig.~6) and substructure at the high frequencies cause
some departures from a $\lambda^2$ law.  There is significant
depolarization (9.5\% at 15.3 GHz to 3.5\% at 8.2 GHz at $\sim$ 1.5
mas resolution) in the southern hot spot.  This depolarization, and
that seen between 5 and 22 GHz in a larger beam, are consistent with
beamwidth depolarization by the observed RM gradient.

It is worth noting that the RM decreases with increasing distance from
the center of the lobe, i.e., the gradient slopes down in the
direction of the edge of the lobe.  This indicates that the
density and/or the field strength decreases towards the edge.
This situation is reminiscent of that in M87 where no polarization
is detected in the bright inner few parsecs \citep{zav02}, and is 
most naturally explained by a radial falloff in the density.
Polarization is not expected from the counterjet in 3C\,84, both because
it is fainter, and because it is behind a denser Faraday screen.

The RM corrected plot of the projected magnetic field orientation of
the linearly polarized flux at 15 GHz is shown in Fig~5.  The
intrinsic magnetic field in the southern jet component S1 appears to
be predominantly perpendicular to the jet axis, as expected if the
field is enhanced by compression.  The polarization angle of the core
is uncertain given that the RM of the core is not well determined
and is likely to be well in excess of 10000 \radm.

We note that \cite{hom04} found strong circular polarization in the
central parsec of 3C\,84, reaching +3\% at 15 GHz.  They speculate
that the circular polarization may be produced by Faraday conversion
of linear to circular polarization.  No circular polarization was
detected by \cite{hom04} from the bright southern jet at 15 or 22 GHz, and
the linear polarization from the central parsec was found to be 
less than 1\%.

\section{Discussion}

\subsection{Magnetic Fields in Perseus}\label{discussion-jet}

For a refractive medium in the presence of magnetic fields the 
intrinsic polarization angle,
$\chi_0$, is observed as $\chi$ such that
\begin{equation}
\chi = \chi_0 + RM\lambda^2
\end{equation}
where $\lambda$ is the observed wavelength.  The rotation measure, RM,
is related to the electron density, $n_e$, the net line of sight
magnetic field in the environment, $B_\|$, and the path length, $dl$, through the plasma,
by the equation
\begin{equation}
RM = 812\int n_e B_\| dl \quad \mbox{rad m$^{-2}$}
\end{equation}
where units are in cm$^{-3}$, $\mu$G, and kiloparsecs.  
Our best estimate from \S \ref{XrayObs} for $n_e$ is 0.3 cm$^{-3}$.
We assume a path length of 2 kpc, which probes the highest density
gas in the cluster, and is typical of RM scale sizes in other
cooling core clusters \citep{car02}.
Assuming a constant magnetic field orientation, we find a magnetic 
field strength of 15 $\mu$G.
This is only the component along the line-of-sight, so correcting 
by a factor of $\sqrt{3}$ we estimate a field strength of 25 $\mu$G.

Field strengths calculated with these parameters can be compared to
the strength of a magnetic field that has the same pressure as a
gas of the same n$_e$ and a temperature of 5 $\times$ 10$^7$~K using
\begin{equation}
{B^2 \over {8\pi}} = {2 n_e k T \mbox{.}}
\end{equation}

In the central ($r < 2$ kpc) region of the Perseus cluster we find
this gives 300 $\mu$G, so the magnetic pressure from the 
estimated field strength of 25 $\mu$G is two orders of magnitude 
less than the 
thermal pressure ($\sim 4~\times~ 10^{-9}$ dyn
cm$^{-2}$).  This result is similar to that found in other cooling
core clusters.  

A difficulty with producing the RMs in 3C\,84 in the
ICM is that the observed gradient of 10\% of the RM on scales of
$\sim$1 pc is hard to reconcile with fields organized on kpc scales.

\subsection{The high-velocity system}

Emission- and absorption-line studies have highlighted the existence of 
a high-velocity system (at $\sim$ 8200 km s$^{-1}$) at approximately the 
same position on the sky as NGC 1275 (which has a systemic velocity of 
$\sim 5200$ km s$^{-1}$). This system is likely to be associated with
a spiral galaxy falling into the Perseus cluster at 3000 km s$^{-1}$.
Recently \cite{gill04} have shown through the study of the X-ray absorption
that this system is not interacting with the body of NGC 1275 and that 
they are separated by at least 57 kpc.

Can this spiral galaxy be responsible for the observed rotation measures? 
The hot component of its ISM is not likely to produce the observed
RMs, as the required density (assuming a $B_\|$ of 3 $\mu$G, a
path length of 10 kpc and a temperature of $10^5$ K) is 0.3 cm$^{-3}$, 
about two orders of magnitudes larger than one would infer from a 
simple pressure equilibrium argument with the neutral component of 
the ISM, whose density and temperature were estimated by \cite{mom02}.
The possibility that a very compact H II region belonging to the
high-velocity system might be responsible for the observed RMs cannot 
be completely excluded, even though the probability of finding such
a region directly along our line-of-sight to the core of NGC 1275 is
very small.

\subsection{The ionized filaments}

The filamentary structure of ionized gas associated with NGC 1275,
instead, might well produce the observed RMs: the H$\alpha$
observations performed by \cite{con01} showed the presence of
unresolved features with size less than $\sim$230 pc and lower bound
electron densities of $\ge~10$ cm$^{-3}$. Densities measured in the
[SII] lines in the central kiloparsecs are 270 cm$^{-3}$
\citep{joh88,hec89}, and the optical filaments are thought to be in
equipartition with the ICM with pressures of $\sim 1~\times~ 10^{-9}$
dyn cm$^{-2}$.  Comparing the central surface brightness in H$\beta$
with emission from gas at the X-ray pressure \citep{fab03} near the
center of the cluster gives a depth of 0.06 pc for a temperature of
10$^4$ K and a uniform covering fraction, $f$.  If the filaments are
more filamentary than sheet-like, then the depth of the H$\beta$
increases in proportion to 1/$f$, but it becomes difficult with a
small covering factor to produce coherent rotation measures across the
radio source.  Also, if the surface brightness of the H$\beta$ line
rises close to the nucleus where it is unresolved, then this can
increase the estimate of the depth.  Observations in the Pa$\alpha$
line \citep{wil05} indicate that the surface brightness does rise by
one to two orders of magnitude within the inner 150 pc. This increases
the above estimate for the depth of the filaments from the H$\beta$
line to at least 1 parsec.

The small size of the ionized filaments could explain the 10\%
gradient in the RM on scales of a parsec.  Assuming a constant
gradient we might use this to estimate a scale size for the RMs of
$\sim$10 parsecs.  Assuming that the scale size along the
line-of-sight is similar to that in the plane of the sky leads to an
estimate for the Faraday depth of $\sim$10 pc. 

With the above density and path length (10 pc) and the observed RM we obtain a
lower limit for $B_\|$ of 4 $\mu$G, and hence a magnetic pressure of
$\sim 2~\times~ 10^{-12}$ dyn cm$^{-2}$, nearly three orders of
magnitude below the thermal pressure.  If the filaments are as small
as 1 pc (with a uniform covering factor, $f=1$), then the field strength
required to produce the observed RMs is $B \sim 50 \mu$G, but the magnetic
pressure is still less than the thermal pressure by a factor of 
$\sim$4. 

It is also possible that the radio source is interacting with the
ionized gas and that the densities are enhanced at the southern edge
of the expanding source, which appears reminiscent of a bow-shock.
Locally higher densities would further reduce the strength of the
magnetic fields required to produce the observed RMs.  Similar
magnitude RMs have been detected in a few other well studied radio
galaxies (e.g., M87, 3C120 \cite{zav02}), and attributed to ionized
gas in close proximity to the radio emission.  Baum et al. (2005)
\nocite{bau05} have shown that there is no significant column density of
more neutral gas along the line of sight to the nucleus from the lack
of any strong Ly$\alpha$ absorption.

We note that the observed decrease in fractional polarization 
with wavelength can be attributed to gradients in the Faraday
screen.

\section{Conclusions}

We find substantial Faraday Rotation Measures of $\sim$7000 \radm\ toward
3C\,84.  RMs as large or larger than this have been suspected for
some time due to the low observed polarization from this bright 
radio galaxy.  The Faraday screen is most likely to be associated
with the ionized gas that also produces spectacular filaments 
of H$\alpha$ emission in the Perseus cluster.  This gas may well
have magnetic fields organized on small enough scales ($<$ 10 pc),
to produce the observed gradient in the RM.  

Our current measurements provide only a few very closely spaced
lines-of-sight through the cluster.  To establish the scale size over
which the magnetic fields are organized, and to look for correlations
with the cavities seen in the X-ray emission, we would like to be able
to sample the RM distribution in the Perseus cluster on scales out to
many 10s kiloparsecs.  There is fairly strong radio emission on these
scales, but detecting the polarization is challenging.  To determine
RMs in larger regions we have to have greater sensitivity to extended
emission than the VLBA provides.  But increasing the beam size means
more susceptibility to RM gradients within the beam.  It also means
that the flux within the central resolution element goes up (quite
dramatically when components C and S1 merge) so the required
dynamic range increases as well.  Errors in the leakage calibration
are typically $\sim$0.5\% for individual antennas and will average 
out over an array by $\sim N^{1/2}$ where N is the number of elements
in the array; errors are further reduced away from the image center 
by an additional factor of $N^{1/2}$ \citep{rob94}.  These errors 
scale with the total intensity so that
a point source of 10 Jy produces a linear polarization noise floor
of $\sim$5 mJy/beam for the VLBA and $\sim$2 mJy/beam for the VLA.  

An array like the proposed EVLA phase 2 with 35 antennas, excellent
sensitivity, and sub-arcsecond resolution could have a good chance of
detecting polarization from a larger region across 3C\,84 at high
frequencies.  The EVLA phase 1 currently under construction has less
resolution than desired, but if the errors in the leakage terms can be
reduced below 0.5\%, then it might be possible for some regions where
the gradient happens to be low to be measured.
This will also become easier in time so long as 3C\,84 continues to fade 
at centimeter wavelengths \citep{all03}.

Future VLBA observations of 3C\,84 to look for changes in the 
RM distribution with time could provide additional information about the
nature of the Faraday screen. 

\acknowledgments We are grateful to the referee, Daniel Homan, for
insightful suggestions.  GBT acknowledges support for this work from the
National Aeronautics and Space Administration through Chandra Award
Number GO4-5134A issued by the Chandra X-ray Observatory Center, which
is operated by the Smithsonian Astrophysical Observatory for and on
behalf of the National Aeronautics and Space Administration under
contract NAS8-03060.  NEG gratefully acknowledges support from the
NRAO Graduate Summer Student Research Assistantship.  
This research has made use of the
NASA/IPAC Extragalactic Database (NED) which is operated by the Jet
Propulsion Laboratory, Caltech, under contract with NASA.  The
National Radio Astronomy Observatory is a facility of the National
Science Foundation operated under a cooperative agreement by
Associated Universities, Inc.

\clearpage

\begin{figure}[h!]
\psfig{figure=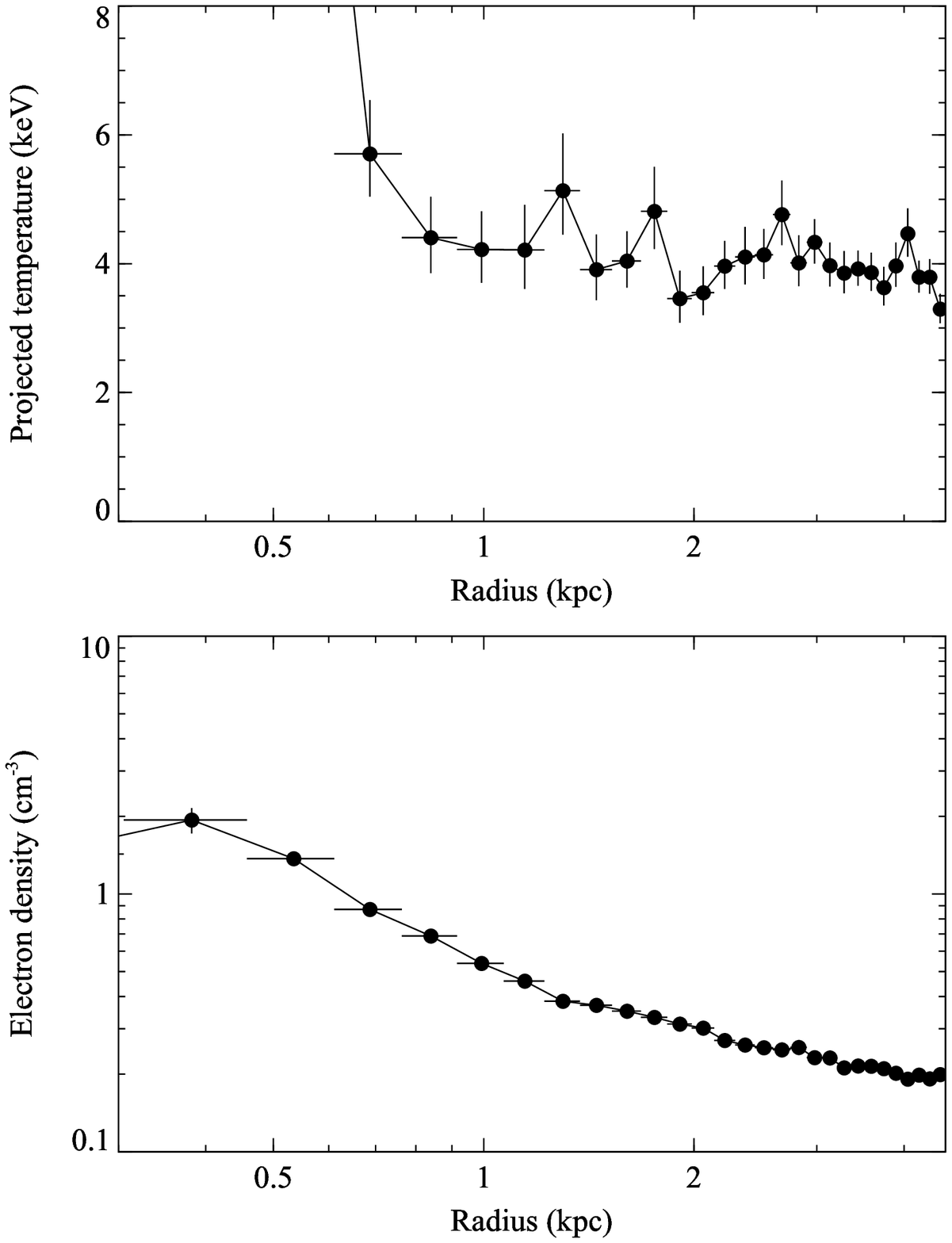,height=6.5in}
\vskip-0.5cm
{\caption[]{\label{profile} {\small
The temperature and density profile in the center of the Perseus 
cluster derived from a long Chandra observation \citep{fab06}.
}}}
\end{figure}
\clearpage

\begin{figure}[h!]
\psfig{figure=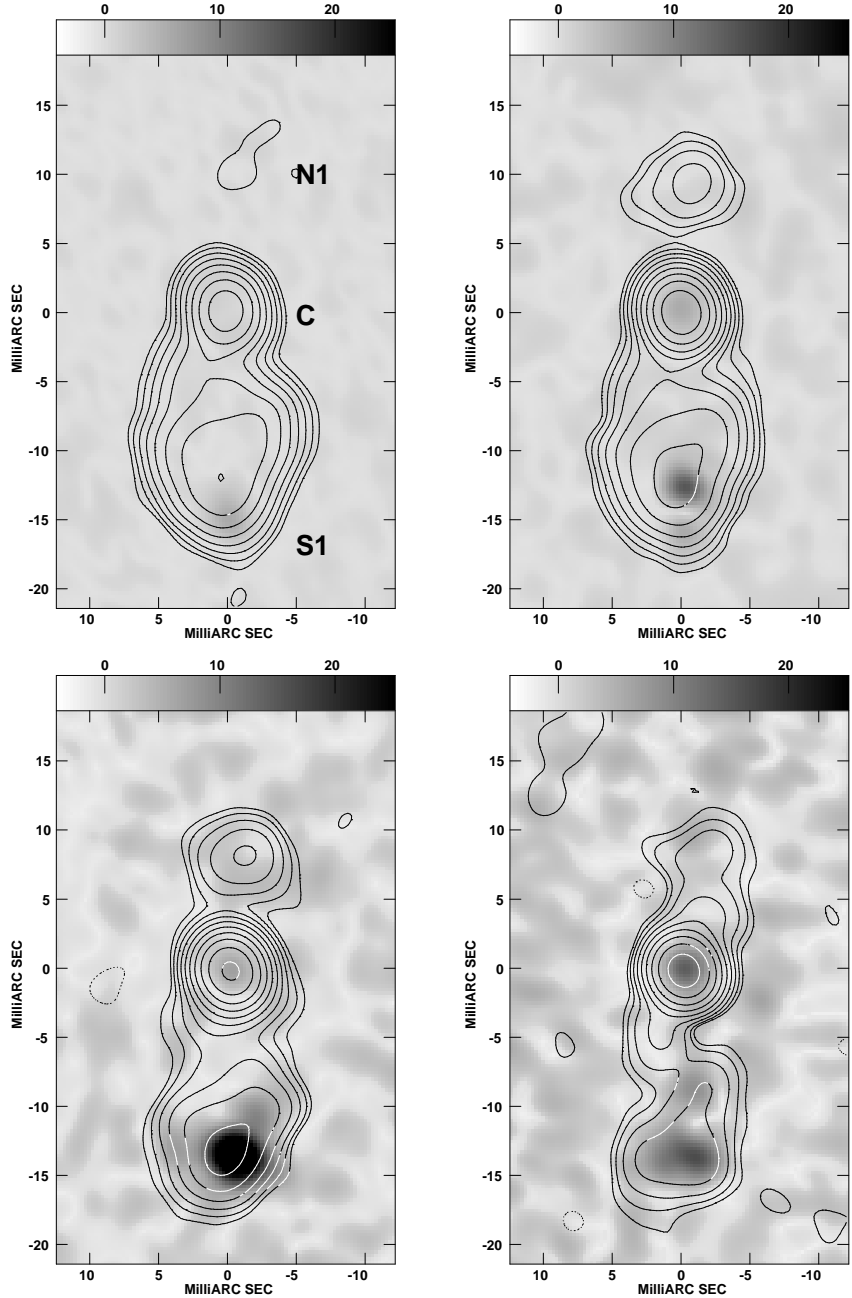,height=8.5in}
\vskip-2cm
{\caption[]{\label{mosaic} {\small
VLBA observations of 3C\,84 at 5, 8, 15, and 22 GHz at a fixed 
angular resolution of 2.75 mas.  The greyscale represents linearly polarized
intensity ranging from -4 to 25 mJy/beam in all panels.  Contour
levels begin at 15 mJy/beam and increase by factors of 2.  
Coordinates are relative to the 
VLBA pointing center at J2000 R.A. 03 19 48.16, Dec. 41 30 42.1.
}}}
\end{figure}
\clearpage

\begin{figure}[h!]
\psfig{figure=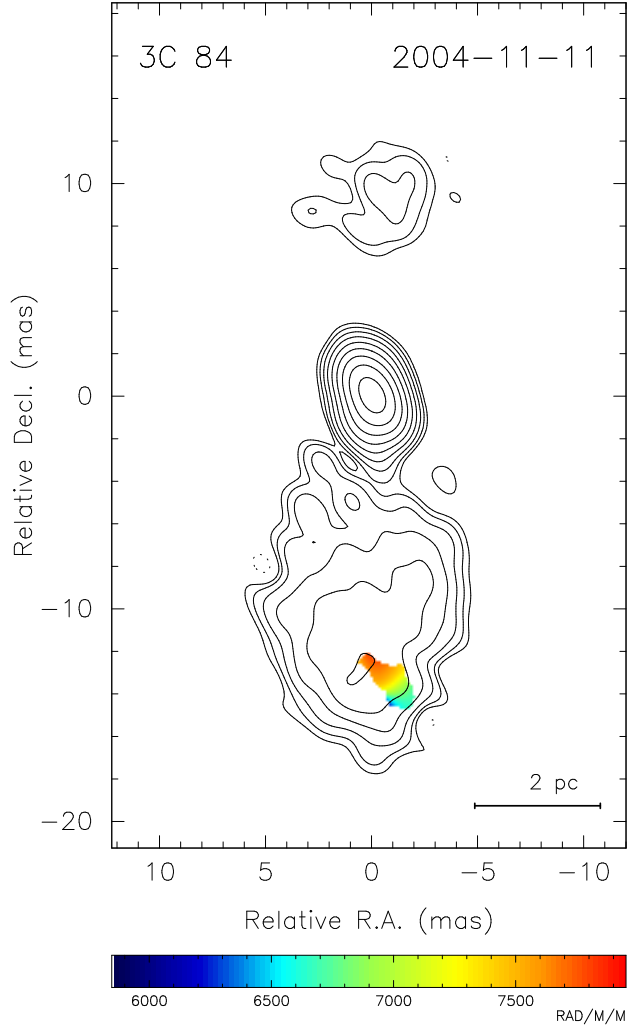,height=6.6in}
{\caption[]{\label{rm} {\small
Rotation measure image of 3C\,84 at 1.8 $\times$ 1.3 mas$^2$ resolution 
in position angle 30$^\circ$.  Contour
levels from the 8 GHz total intensity image begin at 15 mJy/beam and increase by factors of 2.  
}}}
\end{figure}
\clearpage

\begin{figure}[h!]
\psfig{figure=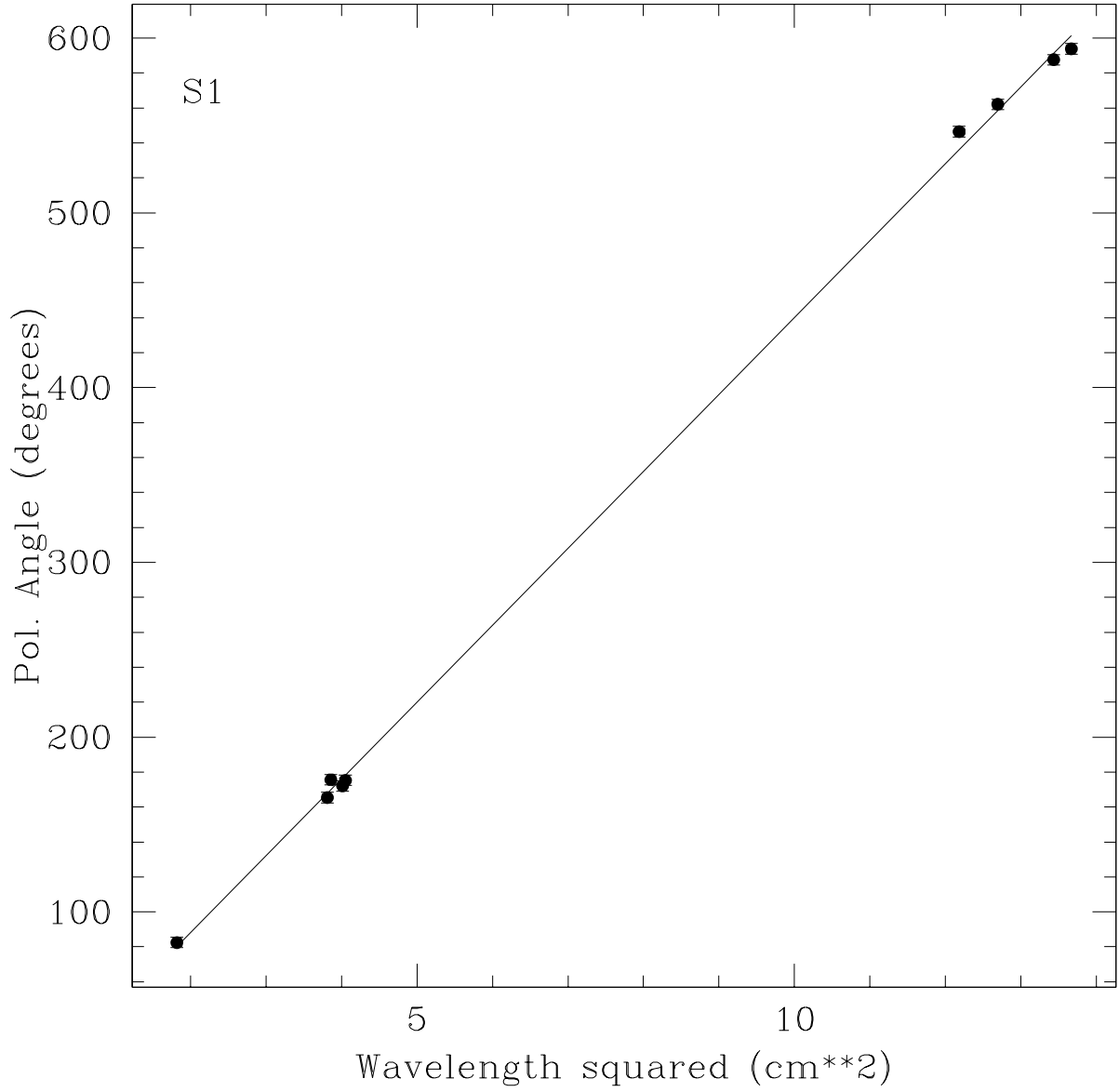,height=6.6in}
{\caption[]{\label{s1} {\small
Position angle versus $\lambda^2$ for the southern component.  The
fit gives the Faraday rotation measure of 7680 $\pm$ 64 \radm.
}}}
\end{figure}
\clearpage

\begin{figure}[h!]
\psfig{figure=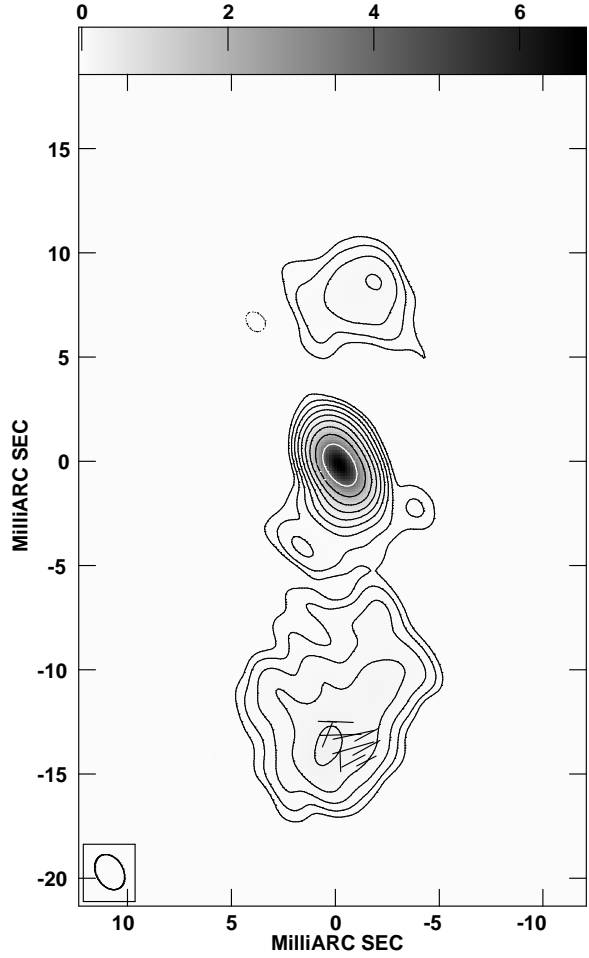,height=6.6in}
{\caption[]{\label{bvec} {\small
The RM corrected magnetic field (B) vectors at 15 GHz overlaid
on a total intensity image at 1.8 $\times$ 1.3 mas$^2$ resolution 
in position angle 30$^\circ$.  The
length of the vectors is proportional to polarized flux density.
Contour levels begin at 15 mJy/beam and increase by factors of 2.  
}}}
\end{figure}
\clearpage

\begin{figure}[h!]
\psfig{figure=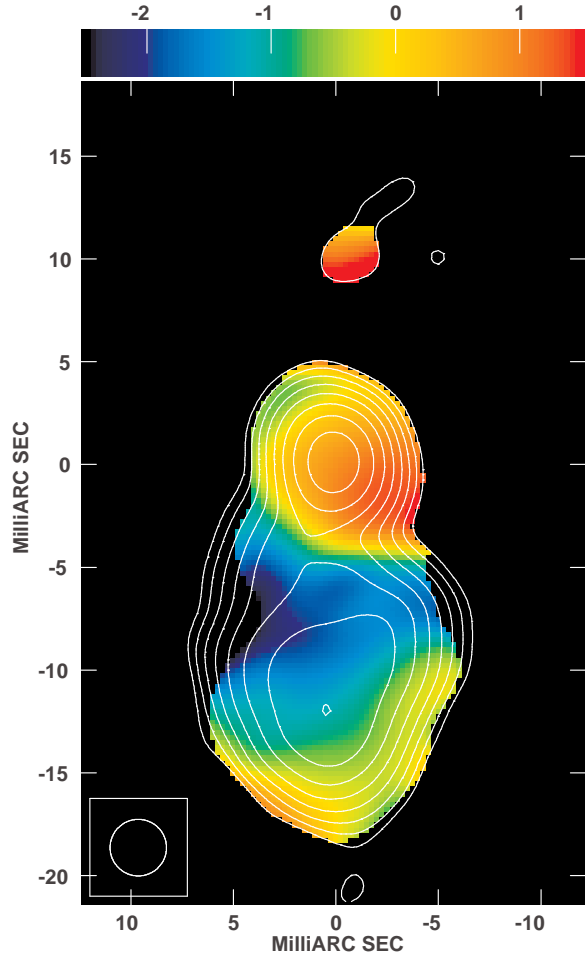,height=6.6in}
{\caption[]{\label{spix} {\small
The spectral index image, defined $S_\nu \propto \nu^\alpha$, between 5 and 15 GHz overlaid on the
5 GHz total intensity image at 2.75 mas resolution.  
Contour levels begin at 15 mJy/beam and increase by factors of 2.  
}}}
\end{figure}
\clearpage


\clearpage

\begin{deluxetable}{lcccccc}
\tabletypesize{\footnotesize}
\tablewidth{0pt}
\tablecolumns{6}
\tablecaption{Observational Parameters\label{tab1}}
\tablehead{\colhead{Source} & \colhead{Date} & \colhead{Freq.} &
  \colhead{Time} & \colhead{Bandwidth}  & \colhead{Peak} & \colhead{rms}\\ 
  \colhead{} & \colhead{} & \colhead {(GHz)} & \colhead{(min)} &
  \colhead{(MHz)} & \colhead{(mJy/beam)} & \colhead{(mJy/beam)}\\
  \colhead{(1)} & \colhead{(2)} & \colhead{(3)} & \colhead{(4)} &
  \colhead{(5)} & \colhead{(6)} & \colhead{(7)}}
\startdata
3C84 & 20041026 & 4.6 & 11 & 16 & 3200 & 0.52 \\
 & 20041026 & 5.0 & 11 & 16 & 3560  & 0.96 \\
 & 20041026 & 8.2 & 10 & 16 & 7390 & 1.20 \\
 & 20041026 & 8.5 & 10 & 16 & 7620 & 1.21 \\
 & 20041026 & 14.9 & 10 & 16 & 8960 & 2.27 \\
 & 20041026 & 15.3 & 10 & 16 & 8910 & 2.25 \\
 & 20041026 & 22.2 & 10 & 32 & 5990 & 7.63 \\
3C84 & 20041111 & 4.6 & 13 & 16 & 3120 & 0.31 \\
 & 20041111 & 5.0 & 13 & 16 & 3520 & 0.39 \\
 & 20041111 & 8.2 & 12 & 16 & 7430 & 0.56 \\
 & 20041111 & 8.5 & 12 & 16 & 7600 & 0.36 \\
 & 20041111 & 14.9 & 13 & 16 & 8750 & 1.38 \\
 & 20041111 & 15.3 & 13 & 16 & 8750 & 1.37 \\
 & 20041111 & 22.2 & 14 & 32 & 5940 & 1.36 \\
\enddata
\tablenotetext{*}{
Notes - (1) J2000 source name; (2) Date of observation; (3) Frequency
in GHz; (4) Integration time in minutes; (5) Bandwidth in MHz; (6)
peak in the total intensity image from the average of two 
adjacent IFs; (7) rms
noise off source from the Stokes Q image in mJy/beam measured from the 
matching resolution images.  Noise in Stokes I, Q and U images are similar.}
\end{deluxetable}
\clearpage

\begin{center}
Table 2. Source Properties \\
\smallskip
\begin{tabular}{l c c}
\hline
\hline
Property & 3C\,84 \\
\hline
\noalign{\vskip2pt}
core RA (J2000) & 03$^h$19$^m$48\rlap{$^s$}{.\,}1601 \\ 
\phantom{core}Dec. (J2000) &  41\arcdeg 30\arcmin 42\rlap{\arcsec}{.\,}104 \\
\phantom{core}Gal. long. ($l$) & 150.58\dg \\
\phantom{core}Gal. lat. \phantom{   }($b$) & $-$13.26\dg  \\
radial velocity & 5264 $\pm$ 11 km s$^{-1}$ \\
distance from cluster center &  0.0 Mpc \\
luminosity distance & 75.4 Mpc \\
core flux density (5 GHz) & 3.1 $\pm$ 0.16 Jy \\
core power (5 GHz) & 2.1 $\times$ 10$^{24}$ W Hz$^{-1}$ \\
largest angular size & 200\arcsec \\
largest physical size & 70 kpc \\
total flux density (5 GHz) & 23.3 Jy \\
total power (5 GHz) & 1.6 $\times$ 10$^{25}$ W Hz$^{-1}$ \\
\hline
\end{tabular}
\end{center}

\clearpage

\end{document}